\documentclass[aps,pre,twocolumn,amsfonts,groupedaddress]{revtex4}

\begin{document}

\title{ Comment on ``Entropy and Wigner function'' }

\author{Joachim J. W{\l}odarz}

\email{jjw@tkemi2.klb.dtu.dk}

\affiliation{ Chemical Physics, Department of Chemistry,
                 Technical University of Denmark,
                 DTU--207, DK--2800 Lyngby, Denmark }

\affiliation{ Department of Theoretical Chemistry, Silesian University,
              Szkolna 9, PL--40006 Katowice, Poland }

\date{\today}

\begin{abstract}
In a recent paper [Phys. Rev. E {\bf 62}, 4556 (2000)] Manfredi and Feix 
proposed an alternative definition of quantum entropy based on Wigner 
phase-space distribution functions and discussed its properties. They
proposed also some simple rules to construct positive-definite Wigner
functions which are crucial in their discussion. We show, however, that
the main results obtained  and their discussion are incorrect.
\end{abstract}

\pacs{ 05.30.-d, 03.65.-w }

\maketitle

Entropy is without any doubt one of the most important physical concepts,
not only in thermodynamics and statistical mechanics, but also in various
areas of quantum theory. It remains to be also one of the most mysterious 
quantities, especially outside the traditional phenomenological 
thermodynamics.

In a recent paper \cite{PRE_62_4665} Manfredi and Feix proposed an 
alternative definition of quantum entropy with quadratic dependence on 
the Wigner distribution function $W(x,p)$:
\begin{equation} \label{MFe}
S_2 = 1 - (2\pi\hbar)^D \int_{\Gamma}dxdp\; (W(x,p))^2
\end{equation}
The main advantage of this expression is that it is an exact Weyl 
transform of the linear entropy:
\begin{equation} \label{MFe2}
S_2 = 1 - {\mathrm Tr}(\widehat{\rho}^2)
\end{equation}
and therefore  all properties of the linear entropy could be easily 
extended to Eq.(\ref{MFe}).

For a quantum state {\em per se}, i.e., without any measurements being
involved, represented through a density operator $\widehat{\rho}$, the 
(dimensionless) quantum entropy may be defined after von~Neumann 
\cite{GN_1927_279} as:
\begin{equation} \label{VNe}
S = - {\mathrm Tr}(\widehat{\rho}\ln \widehat{\rho} )
\end{equation}
The presence of the $\ln \widehat{\rho}$ term precludes an easy 
translation of Eq.(\ref{VNe}) to, e.g., the Weyl-Wigner-Moyal 
phase-space picture of quantum mechanics.

The classical limit of the von Neumann entropy is given by the Wehrl
entropy \cite{RepMathPhys_16_353,JMP_25_1507}:
\begin{equation} \label{We}
S_{cl} = - \int_{\Gamma}dpdq\; 
                       \rho_H(p,q;\kappa)) \ln \rho_H(p,q;\kappa)
\end{equation}
where $\rho_H$ denotes the respective Husimi phase-space distribution 
function parametrized with some arbitrary positive constant $\kappa$.
The Wehrl entropy is possibly the closest quantity to the classical
Gibbs entropy (Eq.(4) in \cite{PRE_62_4665}), but they are obviously 
very different from each other.
Therefore, the discussion of the classical limit of the von~Neumann
entropy presented in Sec. I of \cite{PRE_62_4665}, although appealing,
must be considered as incorrect.

The linear entropy Eq.(\ref{MFe2}), as well as the term
${\mathrm Tr}(\widehat{\rho}^2)$ alone, is obviously a measure of the 
purity of the involved quantum state, therefore the  opinion 
(\cite{PRE_62_4665}, p. 4666) that $\int W^2 dxdp$  
``has no obvious physical meaning'' seems to be surprising.

Discussing the properties of the quantum entropy, the authors have found
that the Wigner distribution function which maximizes the entropy is defined 
as follows (\cite{PRE_62_4665}, p. 4668):
`` $W={\mathrm const}=\Omega^{-1}$ within a phase space region of volume 
(area) equal to $\Omega$, and $W=0$ elsewhere'', 
i.e.,  $W$ is simply a characteristic function of this phase space area
normalized to unity.
It could be shown that such a phase-space function, and in general {\em any}
compactly supported phase-space function {\em cannot} be regarded as a valid 
Wigner distribution function 
(cf. \cite{JJW_PLA133_459,DL_PhysicaA182_643,JFAA4_723}
and references therein).

In Sec. IV (\cite{PRE_62_4665}, p. 4669) the authors discuss extensively the
Gaussian smoothing of Wigner distribution functions. A claim is made that
the result of Gaussian smoothing of any Wigner distribution function 
``... is still an admissible Wigner function''. The admissibility is then
defined in exactly the same way as the necessary and sufficient
conditions for a phase-space function to be a Wigner distribution
function (Ref.~4 in \cite{PRE_62_4665}, see also \cite{PRA_34_1}).
In other words, any admissible phase-space function have to be a genuine
Wigner distribution function and {\em vice versa}.

It is well known (cf., e.g., \cite{PRep_259_147}) that the Gaussian smoothed
Wigner phase-space distribution functions are Husimi functions, which  are
associated with the (generalized) antinormal operator ordering. Also, Husimi
functions does not give proper marginal densities in the general case.
In contrast, the Weyl-Wigner representation employs a different, symmetrized 
operator ordering, and the Wigner distribution functions give always  
proper marginal densities both in position and momentum spaces.
Therefore, the claim that Gaussian smoothing of any Wigner distribution 
function gives always a non-negative Wigner distribution function 
{\em cannot} be true as inconsistent with well known results, and the offered 
proof must be flawed.

Indeed, the key point in the authors' reasoning is the substitution of 
$W(x,p)$, corresponding to $\psi(x)$, by $W_1(x,p)=W(-x,-p)$ 
(corresponding to $\psi(-x)$) made in Eq.(33) of \cite{PRE_62_4665}. 
This substitution is {\em not} legitimate in the general case, because 
only $W(x,p)$ obeying $W(x,p)=W(-x,-p)$ could be allowed here.

To demonstrate that explicitly, let us repeat here orderly the steps 
performed in \cite{PRE_62_4665}, starting from the l.h.s of Eq.~(32) with
$\overline{W}(x,p)$ taken from Eq.~(30):
\begin{eqnarray}  \label{X}
&&\int\!dxdp\;\overline{W}(x,p) F(x,p)                               \\
&&=\int\!dxdp F(x,p) \int\!dx'dp'\;W(x',p') K(x-x',p-p') \nonumber   \\
&&=\int\!dxdp F(x,p) \int\!dx''dp''\;W(x-x'',p-p'') K(x'',p'') \nonumber \\
&&=\int\!dx''dp'' K(x'',p'') \int\!dxdp\;W(x-x'',p-p'') F(x,p) \nonumber
\end{eqnarray}
The last expression Eq.~(\ref{X}) obtained above is equivalent to
\begin{equation} 
\int\!dx''dp'' K(x'',p'') \int\!dxdp\;W(x''-x,p''-p) F(x,p)
\end{equation}
containing the desired convolution $W \circ F(x'',p'')$,
{\em if and only if}:
\begin{equation}
W(x,p) \equiv W(-x,-p)
\end{equation}

The suggested further (\cite{PRE_62_4665}, p. 4670) way:
``to construct a phase-space distribution function which is both
positive and admissible [...] by smoothing an arbitrary (but positive)
function of phase space variables, again with a Gaussian kernel. ...'' 
is also incorrect. This should be not a surprise even at a first glance, 
because an arbitrary, but non-negative phase-space function will be 
almost always {\em not} a sufficiently regular one. 
A simple counterexample may be furnished here by the function:
\begin{equation}
f_{a}(x,p)=\exp\left\{ a(x^2+p^2)\right\}, \; a > a_0 > 0
\end{equation}
where the parameter $a$ is sufficiently large (the threshold value $a_0$
depends on the smoothing Gaussian). This everywhere positive phase-space 
function {\em cannot} be smoothed with a Gaussian, because the 
respective integrals are divergent.

Finally, the procedure of taking the classical limit of  Eq.(\ref{MFe}) 
by ``keeping the Wigner function fixed, and letting Planck's constant go 
to zero'', discussed in Sec. VI (\cite{PRE_62_4665}, p. 4673)
cannot be so simple, due to the explicit dependence of the Wigner 
distribution function on the Planck's constant. An excellent and detailed
discussion of these and related problems may be found in 
\cite{PTRSLA_287_237}. However, for pure states we have:
\begin{equation} \label{Wsq}
\int_{\Gamma}dxdp\; W(x,p)^2 = (2\pi\hbar)^{-D}
\end{equation}
which in combination with Eq.(\ref{MFe}) gives always $S_2=0$ rather than
$S_2=1$ (indicated as the ``unpleasant property'' of $S_2$ in 
\cite{PRE_62_4665}).

\bigskip

\begin{acknowledgments}
It is a pleasure to thank Professor Jens Peder Dahl for his extremely
warm hospitality. The support received from the Danish National Research
Council is also acknowledged.
\end{acknowledgments}



\bigskip

\begin{center}
{\large\bf Addendum}
\end{center}

\bigskip

This Comment has been rejected by the PRE Editorial Board as not  
{\em ``[...]~helpful in straightening the matter~[...]''}
(quoted from the Editorial Board Member report).

Whatever it is supposed to mean, let me only stress here once again 
the most important issues raised in my Comment:

\bigskip

\begin{itemize} 

\item Gaussian-smoothed Wigner distribution functions are in general
      {\em not} Wigner distribution functions themselves. 

\item {\em Arbitrary} positive functions of phase-space variables
      {\em cannot} be in general converted to Wigner distribution functions
      via Gaussian smoothing.
      
\end{itemize}

\bigskip

These issues are crucial for the criticized paper, where quantum entropy is
defined as a state functional in the Weyl-Wigner-Moyal picture
(cf. Eq.(8) in \cite{PRE_62_4665}). 
The usage of Gaussian-smoothed Wigner distribution functions instead of
ordinary Wigner distribution functions means here a {\em representation 
switch} to the Husimi picture, with all consequences for the derived results. 
For example, the inequality $S_2[\overline{W}] \geq S_2[W]$
(cf. Eq.(44) in \cite{PRE_62_4665}) does {\em not} indicate that smoothing
generally increases the $S_2$ entropy, because $S_2[\overline{W}]$ and 
$S_2[W]$ are well-defined in {\em distinct} phase-space representations.

\end{document}